\documentstyle[11pt]{llncs}

\begin{document}
\title{The Undecidability of Unification Modulo $\sigma$ Alone}
\author{Gilles Dowek}
\date{}
\institute{\'Ecole polytechnique and INRIA\\
LIX, \'Ecole polytechnique,
91128 Palaiseau Cedex, France.\\
{\tt gilles.dowek@polytechnique.edu}}
\maketitle

\begin{definition}
A {\it second order context} is a context in which all types are of order
at most 2.

In $\lambda$-calculus, a unification problem $a = b$ is a {\it second
order unification problem} if its context is second order, the common
type of $a$ and $b$ is atomic and all the free variables of $a$ and
$b$ have a type of order at most 2.
\end{definition}

\begin{proposition}
There is no algorithm that decides if a second order unification
problem has a solution or not.
\end{proposition}

\proof{See \cite{Goldfarb}.}

\begin{definition}
A $\lambda \sigma$-term is said to be {\it simple} if every of its
subterms of the form $X[s]$ is such that $s = \uparrow^{n}$ for some
$n$.
A substitution is said to be {\it simple} if for every variable $X$,
$\theta X$ is simple.
\end{definition}

\begin{proposition}
A problem $a = b$ has a solution, if and only if $a_{F} = b_{F}$ has a
solution in $\lambda$-calculus, if and only if $a_{F} = b_{F}$ has a
solution in the image of $F$, if and only if $a_{F} = b_{F}$ has a
simple solution.
\end{proposition}

\proof{See \cite{DHK}.}

\begin{proposition}
Let $\sigma$ be the substitution mapping every variable $X$ of sort
$(\Gamma, A_{1} \rightarrow ... \rightarrow A_{n} \rightarrow B)$ occurring in $a = b$ to the
term $\lambda ... \lambda Y$ where $Y$ is a new variable of sort
$(A_{1} ... A_{n}. \Gamma, B)$.
Let $\tilde{a}$ be the normal form of $\sigma a_{F}$ and 
$\tilde{b}$ be the normal form of $\sigma b_{F}$.

The problem $a = b$ has a solution if and only if 
$\tilde{a} = \tilde{b}$ has a simple solution.
\end{proposition}

\proof{The problem $a = b$ has a solution 
if and only if $a_{F} = b_{F}$ has a simple solution.  If $a_{F} =
b_{F}$ has a simple solution mapping $X_{i}$ to $\lambda ...
\lambda c_{i}$ then the substitution mapping $Y_{i}$ to
$c_{i}$ is a simple solution to $\tilde{a} = \tilde{b}$.  Conversely
if $\tilde{a} = \tilde{b}$ has a simple solution mapping $X_{i}$ to
$c_{i}$, the substitution mapping $X_{i}$ to $\lambda ... \lambda
c_{i}$ is a simple solution to problem $a_{F} = b_{F}$.}

\begin{remark}
The context of the equation 
$\tilde{a} = \tilde{b}$ is second order.
The contexts of all the variables occurring in this problem are second
order and their types is atomic. If $X[c_{1}...c_{p}.\uparrow^{n}]$
occurs in this problem then $c_{i}$'s have a first order type.
\end{remark}

\begin{proposition}
Let $a$ be a normal term well typed of atomic type in a second order
context, such that all the variables occurring in $a$ have a second
order context and an atomic type and if
$X[c_{1}...c_{p}.\uparrow^{n}]$ occurs in this problem then $c_{i}$'s
have a first order type.

Let $\theta$ be a simple substitution. 

Then $(\theta a) \downarrow_{\lambda \sigma} = (\theta a)
\downarrow_{\sigma}$.
\end{proposition}

\proof{By induction over the structure of $a$. 

\begin{itemize}
\item
If $a = ({\tt i}~b_{1}~...~b_{n})$ then the types of $b_{i}$'s are
atomic. The result follows by induction.

\item
If $a = X[c_{1}...c_{p}.\uparrow^{n}]$, then let $d = \theta X$
we prove by induction on the structure of $d$ that 
$(d[c_{1}...c_{p}.\uparrow^{n}]) \downarrow_{\lambda \sigma} =
(d[c_{1}...c_{p}.\uparrow^{n}]) \downarrow_{\sigma}$
\begin{itemize}
\item
if $d = (i~e_{1}~...~e_{q})$ then the $e_{i}$'s have an atomic type
and we apply the induction hypothesis,
\item
if $d = X[\uparrow^{r}]$ then the result is trivial.
\end{itemize}
\end{itemize}}

\begin{proposition}
The problem $a = b$ has a solution if and only if $\tilde{a} = \tilde{b}$ has 
a solution for $\sigma$ alone. 

If $a = b$ has a solution, then $\tilde{a} = \tilde{b}$ has a simple
solution, in $\lambda \sigma$ and this it has a solution for $\sigma$
alone.  Converselly if $\tilde{a} = \tilde{b}$ has a solution for
$\sigma$ alone, then it has a solution for $\lambda \sigma$ and thus
$a = b$ has a solution.
\end{proposition}

\begin{corollary}
$\sigma$-unification is undecidable.
\end{corollary}

\begin{remark}
We have reduced the second order unification problem to the
$\sigma$-unification problem. Reducing the full higher order
unification problem seems to be much more difficult. Thus $\sigma$
alone seems to be a formalism that has links with second order languages.
In particular in sorts $A_{1} ... A_{n} \vdash B$, the $A_{i}$'s and $B$
are types and not sorts and thus the ``arrow'' $\vdash$ cannot be nested.

More precisely $\sigma$-calculus seems to be a precise formulation of
``second order languages without $\lambda$'s'' as defined for instance in
\cite{Goldfarb}. We conjecture the decidability of $\sigma$-matching.
\end{remark}

\end{document}